\documentclass {aa}
\usepackage {graphicx}
\def\Teff  {T$_ {\mbox {\scriptsize eff}}$}
\def\logg  {$\log g$}
\def\vt  {$v_ {t}$}

\begin {document}     

\title {{First Stars II. Elemental abundances in the extremely 
metal-poor star 
CS~22949--037}
\subtitle{A diagnostic of early massive supernovae}
\thanks {Based on observations made with the ESO Very Large Telescope 
at Paranal Observatory, Chile (programme ID 165.N-0276(A)).}
}

\author {E. Depagne\inst {1} \and  
V. Hill\inst {1} \and 
M. Spite\inst {1} \and
F. Spite \inst {1} \and
B. Plez\inst {2} \and
T. C. Beers\inst {3} \and
B. Barbuy\inst {4} \and
R. Cayrel\inst {1} \and
J. Andersen\inst {5} \and
P. Bonifacio\inst {6} \and
P. Fran\c{c}ois\inst {1} \and
B. Nordstr\"om \inst {7,5} \and 
F. Primas\inst {8}
} 

\offprints {E. Depagne}

\institute { 
             Observatoire de Paris-Meudon, GEPI,
             F-92195 Meudon Cedex, France\\
   \email {Eric.Depagne@obspm.fr, Vanessa.Hill@obspm.fr,
Roger.Cayrel@obspm.fr}\\
   \email {Monique.Spite@obspm.fr, Francois.Spite@obspm.fr,
Patrick.Francois@obspm.fr }
         \and
             GRAAL, Universit\'e de Montpellier II, F-34095 
Montpellier
             Cedex 05, France\\
   \email {Bertrand.Plez@graal.univ-montp2.fr}
         \and
             Department of Physics \& Astronomy, Michigan State 
University,
             East Lansing, MI 48824, USA\\
   \email {beers@pa.msu.edu}
         \and
    IAG, Universidade de S\~ao Paulo, Departamento de Astronomia, CP
         3386, 01060-970 S\~ao Paulo, Brazil\\
   \email {barbuy@astro.iag.usp.br}
         \and
         Astronomical Observatory, NBIfAFG, Juliane Maries Vej 30,
         DK-2100 Copenhagen, Denmark\\
   \email {ja@astro.ku.dk}
        \and
    Istituto Nazionale di Astrofisica - Osservatorio Astronomico di
Trieste,
    Via G.B. Tiepolo 11, I-34131
             Trieste, Italy\\
   \email {bonifaci@ts.astro.it}
         \and
         Lund Observatory, Box 43, S-221 00 Lund, Sweden\\
   \email {birgitta@astro.lu.se}
        \and 
         European Southern Observatory (ESO),
         Karl-Schwarschild-Str. 2, D-85749 Garching b. M\"unchen, 
Germany\\
   \email {fprimas@eso.org}
}

\date {Received xxx; accepted xxx}

\authorrunning {Depagne et al.}

\titlerunning {CS~22949--037: A diagnostic of early massive 
supernovae}

\abstract {CS~22949--037 is one of the most metal-poor giants known 
([Fe/H]$\approx-4.0$), and it exhibits large overabundances of carbon 
and nitrogen (Norris et al. ).  Using 
VLT-UVES spectra of unprecedented quality, regarding resolution and 
S/N ratio, covering a wide wavelength range (from $\lambda = 350$ to 
$900$ nm), we have determined abundances for 21 elements in this star 
over a wide range of atomic mass.  The major new discovery is an 
exceptionally large oxygen enhancement, [O/Fe] $= 1.97\pm0.1$, as 
measured from the [OI] line at 630.0 nm. We find an enhancement of 
 [N/Fe] of $2.56\pm 0.2$, and a milder one of [C/Fe] $= 1.17\pm$0.1,
 similar to those already reported in the literature. This implies 
 $Z_{\star}=0.01 Z_{\odot}$.
 We also find  carbon isotopic 
ratios $^{12}$C/$^{13}$C$ =4\pm2.0$ and $^{13}$C/$^{14}$N$ =0.03 
^{+0.035}_{-0.015}$, close to the equilibrium value of the CN cycle.  
Lithium is not detected.  Na is strongly enhanced ([Na/Fe] 
$= +2.1 \pm 0.2$),
while S and K are not detected.  The silicon-burning elements 
Cr and Mn are underabundant, while Co and Zn are overabundant 
([Zn/Fe]$=+0.7$).  Zn is measured for the first time in such an 
extremely metal-poor star.  The abundance of the neutron-capture 
elements Sr, Y, and Ba are strongly decreasing with the atomic number of the 
element: [Sr/Fe] $\approx +0.3$, [Y/Fe] $\approx -0.1$, and [Ba/Fe] 
$\approx -0.6$.  Among possible progenitors of CS~22949--037, we 
discuss the pair-instability supernovae.  Such very massive objects 
indeed produce large amounts of oxygen, and have been found to be 
possible sources of primary nitrogen.  However, the predicted odd/even 
effect is too large, and the predicted Zn abundance much too low.  
Other scenarios are also discussed.  In particular, the yields of a 
recent model (Z35Z) from Heger and Woosley are shown to be in fair agreement with 
the observations.  The only discrepant prediction is the very low 
abundance of nitrogen, possibly curable by taking into account other 
effects such as rotationally induced mixing.  Alternatively, the 
absence of lithium in our star, and the values of the isotopic ratios 
$^{12}$C/$^{13}$C and $^{13}$C/$^{14}$N close to the equilibrium value 
of the CN cycle, suggest that the CNO abundances now observed might
have been altered by nuclear processing in the star itself.  A 
30-40$M_{\odot}$ supernova, with fallback, 
seems the most likely progenitor 
for CS~22949--037.  
}

\maketitle
%

\section {Introduction} 

The element abundances in the most metal-poor stars provide the fossil 
record of the earliest nucleosynthesis events in the Galaxy, and hence 
allow the study of Galactic chemical evolution in its earliest phases.  
Moreover, such data can have {\em cosmological} implications: 
Radioactive age determinations (Cowan et al.  \cite {CPK99}; Cayrel et 
al.  \cite {CHB01}) provide an independent lower limit to the age of 
the Universe, and the lithium abundance in metal-poor dwarfs is an 
important constraint on the baryonic content of the Universe 
(Spite \& Spite \cite{SS82}).  The dispersion of the observed 
abundance ratios for a number of elements in extremely metal-poor 
(XMP) stars is very large, reflecting the yields of progenitors of 
different masses, and indicating that each star has been formed from 
material processed by a small number of supernova events -- perhaps 
only one.  Thus, the detailed chemical composition of XMP stars 
provides constraints on the properties of the very first supernovae 
(and hypernovae?)  events in our Galaxy.  Accordingly, the past decade 
has seen a rapidly growing interest in the detailed chemical 
composition of XMP stars.

BPS CS~22947--037, discovered in the HK survey of Beers et al.  
(\cite{BPS92}, \cite{Be99}) is one of the lowest metallicity halo 
giants known ([Fe/H] $\approx-4.0$).  Several studies (McWilliam et 
al.  \cite{MPS95}; Norris, Ryan \& Beers  \cite{NRB01}) have shown that this 
star exhibits a highly unusual chemical composition, characterised by 
exceptionally large enhancements of the lightest $\alpha$ elements: 
Norris et al.  (\cite{NRB01}) found [Mg/Fe] $= +1.2$, [Si/Fe] $= 
+1.0$, [Ca/Fe] $= +0.45$, and even more dramatic values for carbon and 
especially nitrogen: [C/Fe] $= +1.1$ and [N/Fe] $= +2.7$.  In 
contrast, the abundance of the neutron-capture elements was found to 
be rather low: [Ba/Fe] $= -0.77$.

These abundance ratios cannot be explained by classical mixing 
processes between the surface of the star and deep CNO-processed 
layers, or by the enrichment of primordial material by the ejecta of 
standard supernovae.  Canonical supernova and Galactic Chemical 
Evolution (GCE) models (Timmes et al.  \cite{TWW95}) predict that 
[C/Fe] should be solar, and [N/Fe] subsolar (since N is produced as a 
secondary element).  Norris et al.  (\cite{NRB01}) discuss a variety 
of scenarios to explain this abundance pattern, and finally suggest 
that the material from which the star was formed was enriched by the 
ejecta from a massive zero-heavy-element hypernova ($>200 M_{\odot}$), 
which might be able to produce large amounts of primary nitrogen via 
proton capture on dredged-up carbon.

However, no previous analysis has yielded a measurement of the 
abundance of oxygen, an extremely important element, both for the 
precise determination of the nitrogen abundance and as a key 
diagnostic of alternative progenitor scenarios.  We have therefore 
observed CS~22949--037 in the framework of a large, systematic 
programme on XMP stars with the ESO VLT and its UVES spectrograph.  
The extended wavelength coverage (including the red region), superior 
resolution, and S/N of our spectra enable us to determine abundances 
for a large sample of elements with unprecedented accuracy, accounting 
for the abundance anomalies of the star in a self-consistent manner, 
especially in the opacity computation.

This paper presents these new results and discusses their 
astrophysical implications.  In section 2 the observations are sumarized. Section 3 presents a description of themodele atmosphere calculations, and the methods used in the elemental abundance analyses. In section 4 the observed abundance pattern for CS~22949--037 is compared with predictions of recent supernova and hypernova models. 

\begin {table*}[t]
\caption {Log of the UVES observations. $S/N$ refers to the 
signal-to-noise ratio  pixel at 420, 630, and 720 nm in the mean 
spectrum. There are between $6.7$ and $8.3$ pixels per resolution 
element.}
\begin {center}
\begin {tabular}  {lccccccc}
\noalign {\smallskip}\hline
\noalign  {\smallskip}
     &             &            & Exp. & V$_r$  & 420 &  630  & 720\\
date &        UT   & Setting    & time  &km.s$^{-1}$ &  nm &   nm &   
nm\\
\hline
\noalign  {\smallskip}
2000-08-08 & 04:54 & 396 - 573  & 1 h    &-125.68 $\pm$ 0.2    \\
2000-08-08 & 05:57 & 396 - 850  & 1 h \\
2000-08-09 & 05:04 & 396 - 573  & 1 h    &-125.64 $\pm$ 0.2    \\
2000-08-09 & 04:01 & 396 - 850  & 1 h \\
2000-08-11 & 04:27 & 396 - 573  & 1 h    &-125.60 $\pm$ 0.2    \\
2000-08-11 & 05:36 & 396 - 850  & 1 h \\
{\bf Global S/N per pixel (2000/08)}&  &   &  &  &110 &  170   &  
110\\
{\bf Global S/N per resolution element} &  &   &  &  & 285 & 490 & 
320\\
2001-09-06 & 05:06 & 396 - 573  & 1 h    &-125.62 $\pm$ 0.2      \\
{\bf S/N per pixel (2001/09)}&   &        &     &    &40 &  90   &   
- \\
\noalign  {\smallskip}
\hline
\end {tabular}
\end {center}
\label {T-log} 
\end {table*}

\section {Spectroscopic observations} \label{Sec:Spect}

The observations were performed in August 2000 and September 2001 at 
the VLT-UT2 with the high-resolution spectrograph UVES (Dekker et al.  
\cite {D00}).  The spectrograph settings (dichroic mode, central 
wavelength 396 nm in the blue arm, and 573 or 850 nm in the red arm) 
provide almost complete spectral coverage from $\sim$330 to 1000 nm.  
A 1$\arcsec$ entrance slit yielded a resolving power of R$\sim$45,000.

For CS~22949--037 (V $= 14.36$) we accumulated a total integration 
time of 7 hours in the blue, 4 hours in the setting centered at 573 
nm, and 3 hours in the setting centered at 850 nm.  Table \ref{T-log} 
provides the observing log, together with the final $S/N$ obtained at 
three typical wavelengths, and the barycentric radial velocity of the 
star at the time of the observation.

The spectra were reduced using the UVES package within MIDAS, which 
performs bias and inter-order background subtraction (object and 
flat-field), optimal extraction of the object (above sky, rejecting 
cosmic ray hits), division by a flat-field frame extracted with the 
same weighted profile as the object, wavelength calibration and 
rebinning to a constant wavelength, and step and merging of all 
overlapping orders.  The spectra were then co-added and normalised to 
unity in the continuum.  The mean spectrum from August 2000 has been 
used for the abundance analysis.  The spectrum from September 2001 has 
a lower S/N ratio, and has been used only to check for radial velocity 
variations and as a check of the oxygen line profile.

Table \ref{T-log} gives the barycentric radial velocity for each 
spectrum of CS~22949--037.  The zero-point was derived from the 
telluric absorption lines (accurate wavelengths of these lines were 
taken from the GEISA database).  The mean value is $\mathrm{V_r}= 
-125.64\pm0.12$ km.s$^{-1}$ (internal error).  Note that McWilliam et al.  
(\cite{MPS95}) reported a heliocentric velocity $\mathrm{V_r}= 
-126.4\pm0.5$ km.s$^{-1}$ in 1990, while Norris et al.  (\cite{NRB01}) 
obtained $\mathrm{V_r}= -125.7\pm0.2$ km.s$^{-1}$ in September 2000.  
Hence, there is so far no evidence of any significant variation of the 
radial velocity, and thus no indication that CS~22949--037 might be 
part of a binary system.

\section  {Stellar parameters and abundances}

\subsection  {Methods} 

The abundance analysis was performed with the LTE spectral analysis 
code ``turbospectrum'' in conjunction with OSMARCS atmosphere models.  
The OSMARCS models were originally developed by Gustafsson et al.  
(\cite{GBE75}) and later improved by Plez et al.  (\cite{PBN92}), 
Edvardsson et al.  (\cite{EAG93}), and Asplund et al.  (\cite{AGK97}).  
Turbospectrum is described by Alvarez \& Plez (\cite{AP98}) and Hill 
et al.  (\cite{HPC02}), and has recently been further improved by B. 
Plez.

The temperature of the star was estimated from the colour indices 
(Table \ref{param}) using the Alonso et al.  (\cite{AAM99}) 
calibration for giants.  There is good agreement between the 
temperature deduced from $B-V$, $V-R$ and $V-K$.  However, Aoki et al.  
(\cite{ANRBA02}) have shown that, in metal-deficient carbon-enhanced 
stars, the temperature determination from a comparison of broad-band 
colours with temperature scales computed with standard model 
atmospheres is sometimes problematic, because of the strong absorption 
bands from carbon-bearing molecules.  In the case of CS~22949--037, 
although carbon and nitrogen are strongly enhanced relative to iron, 
the molecular bands are in fact never strong because the iron content 
of the star is so extremely low (ten times below that of stars 
analysed by Aoki et al.  \cite{ANRBA00}).  Neither the red CN system nor 
the C$_{2}$ Swan system are visible, and the blue CH and CN bands 
remain weak and hardly affect the blanketing in this region (cf.  
section \ref {sec-CNOalpha}).  Moreover, an independent Balmer-line 
index analysis yields T$_{eff}= 4900$K$\pm125$K, in excellent agreement 
with the result from the colour indices and with the value adopted by 
Norris et al.  (\cite{NRB01}).

\begin {figure}
\resizebox {\hsize}{!}
{\includegraphics {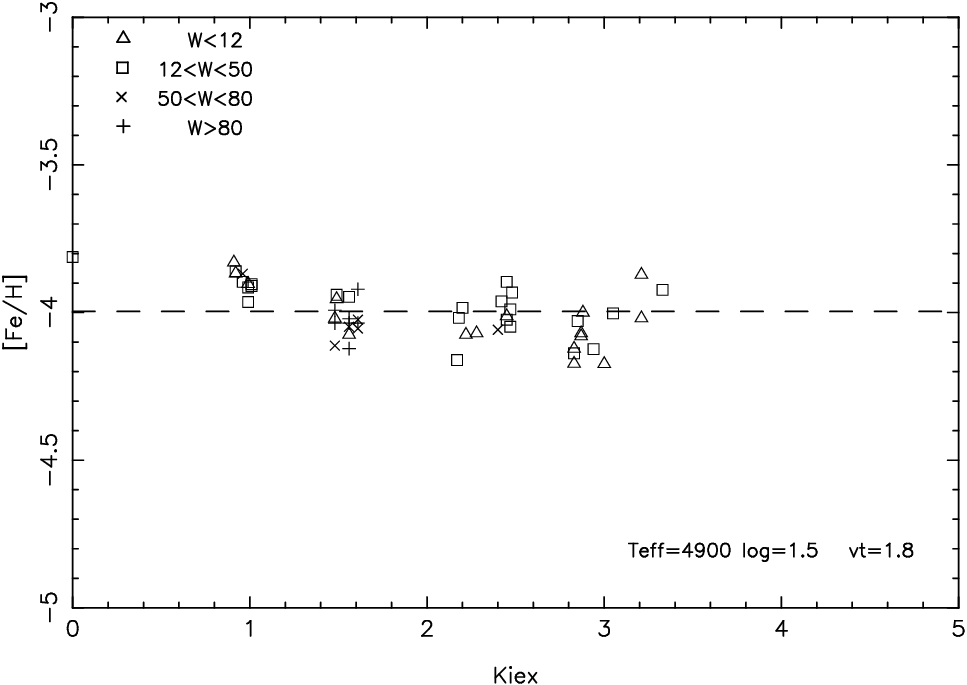}} 
\caption {Iron abundance as a function of excitation potential of the 
line. Symbols indicate different line strengths (in m\AA).} 
\label {kiex}
\end {figure}

With our final adopted temperature, T$_{eff}=4900$K, the abundance 
derived from individual Fe~I lines is almost independent of excitation 
potential (Fig.  \ref{kiex}), at least for excitation potentials 
larger than 1.0 eV. Low-excitation lines are more sensitive to non-LTE 
effects, and the slight overabundance found from the lowest-excitation 
lines is generally explained by this effect (see also Norris et al.  
\cite{NRB01}).

The microturbulent velocity was determined by the requirement that the 
abundances derived from individual Fe I lines be independent of 
equivalent width.  Finally, the surface gravity was determined by 
demanding that lines of the neutral and first-ionized species of Fe 
and Ti yield identical abundances of iron and titanium, respectively.

Table \ref{param} compares the resulting atmospheric parameters to 
those adopted by McWilliam et al.  (\cite {MPS95}) and Norris et al.  
(\cite {NRB01}).

\begin {table}
\caption {Colour indices and adopted stellar parameters for 
CS~22949--037. Temperatures have been computed from the Alonso et al. 
(\cite{AAM99}) relations for E(B-V) $\approx0.03$ (Burstein \& Heiles 
\cite{BH82}). The reddening for different colours has been computed 
following Bessell \& Brett (\cite{BB88}).
 }
\begin {tabular} {lcccc}
\hline
CS~22949--037&      &     &Colour          &        \\
Magnitude or&      &     &corrected for  & \Teff  \\
Colour       &      & Ref &reddening      &(Alonso)\\
\hline
$V$         & 14.36 &  1 &           &             \\
$(B-V)$     & 0.730 &  1 & 0.700     &  4920       \\
$(V-R)_{J}$ & 0.715 &  2 & 0.695     &  4910       \\
$(V-K)$     & 2.298 &  3 & 2.215     &  4880       \\
\hline
\end {tabular}
\begin {tabular} {lcccc}
~\\
\multicolumn{5}{l}{Adopted parameters for CS~22949--037} \\
\Teff       & \logg &  [Fe/H]  & \vt& Ref     \\ 
\hline
4810   & 2.1   &  $-$3.5 & 2.1 & 1          \\
4900   & 1.7   &  $-$3.8 & 2.0 & 4          \\[1.0ex]
{\bf 4900}   &{\bf 1.5}  & {\bf $-$3.9} & {\bf 1.8} & {\bf 5} \\
\hline
\multicolumn{4}{l}{References}\\ 
\multicolumn{4}{l}{1 Beers et al., in preparation}\\ 
\multicolumn{4}{l}{2 McWilliam et al.(1995)}\\ 
\multicolumn{4}{l}{3 Point source catalog, 2MASS survey}  \\
\multicolumn{4}{l}{4 Norris et al. (2001)}  \\
\multicolumn{4}{l}{5 Present investigation} \\ 
\hline 
\end {tabular}
\label {param}
\end {table}

\subsection  {Abundance determinations}\label {sec-abund}

The measured equivalent widths of all the lines are given in the 
Appendix, together with the adopted atomic transition probabilities 
and the logarithmic abundance of the element deduced from each line.  
The error on the equivalent width of the line depends on the $S/N$ 
ratio of the spectrum and thus on the wavelength of the line (Table 
\ref{T-log}).  Following Cayrel (\cite {Ca88}), the error of the 
equivalent width should be about 1 m\AA~ in the blue part of the spectrum and less than 0.5 
m\AA~ in the red .  Since all the lines are weak, 
the error on the abundance of the element depends linearly on the 
error of the equivalent width.  In some cases (where complications due 
to hyper-fine structure, molecular bands, or blends are present) the 
abundance of the element has been determined by a direct fit of the 
computed spectrum to the observations.

Table \ref {abon} lists the derived [Fe/H] and individual elemental 
abundance ratios,  [X/Fe].  The iron abundance measured here is in good 
agreement with the results by McWilliam et al.  (\cite {MPS95}) and 
Norris et al.  (\cite {NRB01}): With an iron abundance ten thousand 
times below that of the Sun, CS~22949--037 is one of the most 
metal-poor stars known today.

As expected for a giant star, the lithium line is not detected.

\begin {table}[h] 
\caption {Individual element abundances.  For each 
element X, column 2 gives the mean abundance $\log \epsilon(X)=\log 
N_{X}/N_{H} + 12$, column 3 the number of lines measured, column 4 the 
standard deviation of the results, and columns 5 and 6 [X/H] and 
[X/Fe], respectively (where [X/H]$ =\log \epsilon(X) - \log 
\epsilon(X)_{\odot}$).} 
\begin {tabular}{lrccrr} 
\hline
  Element &  log $\epsilon(X)$ & n    &  $\sigma$ & [X/H] & [X/Fe]\\
 \hline 
  Fe  I   &   3.51   &   64 &  0.11  & $-$3.99    &       \\
  Fe  II  &   3.56   &    6 &  0.11  & $-$3.94    &       \\
 \hline 
  C (CH)  &   5.72   &      &  synth & $-$2.80    &   +1.17\\
  N (CN)  &   6.52   &      &  synth & $-$1.40    &   +2.57\\
  O  I    &   6.84   &    1 &   -    & $-$1.99    &   +1.98\\
  Na I    &   3.80   &    2 &  0.03  & $-$1.88    &   +2.09\\
  Mg I    &   5.17   &    4 &  0.19  & $-$2.71    &   +1.26\\
  Al I    &   2.34   &    2 &  0.03  & $-$4.13    & $-$0.16\\
  Si I    &   4.05   &    2 &   -    & $-$3.25    &   +0.72\\
  K  I    &$<$0.89   &    2 &   -    & $<-$4.06   &$<-$0.09\\
  S  I    &$<$5.09   &    1 &   -    & $<-$2.12   &$< $+1.78\\	
  Ca I    &   2.73   &   10 &  0.17  & $-$3.63    &   +0.35\\
  Sc II   &$-$0.70   &    5 &  0.14  & $-$3.87    &   +0.10\\
  Ti I    &   1.40   &    8 &  0.09  & $-$3.62    &   +0.35\\
  Ti II   &   1.41   &   21 &  0.15  & $-$3.61    &   +0.36\\
  Cr I    &   1.29   &    5 &  0.10  & $-$4.38    & $-$0.41\\
  Mn I    &   0.61   &    2 &  0.01  & $-$4.78    & $-$0.81\\
  Co I    &   1.28   &    4 &  0.07  & $-$3.64    &   +0.33\\
  Ni I    &   2.19   &    3 &  0.02  & $-$4.06    & $-$0.07\\
  Zn I    &   1.29   &    1 &   -    & $-$3.41    &   +0.70\\
  Sr II   &$-$0.72   &    2 &  0.06  & $-$3.64    &   +0.33\\
  Y  II   &$-$1.80   &    3 &  0.11  & $-$4.04    & $-$0.07\\
  Ba II   &$-$2.42   &    4 &  0.10  & $-$4.55    & $-$0.58\\
  Sm II  &$<-$1.82   &    1 &   -    & $<-$2.83   &$<$+1.14\\
  Eu II  &$<-$3.42   &    1 &   -    & $<-$3.93   &$<$+0.04\\
\hline 
\end {tabular}
\label {abon}
\end {table}

\subsubsection{CNO abundances and the $\mathrm{^{12}C/^{13}C~}$ 
ratio}\label {sec-CNOalpha}

The C and N abundances are based on spectrum synthesis of molecular 
features due to CH and CN. In cool giants, a significant amount of CN 
and CO molecules are formed and thus, in principle, the C, N, and O 
abundances cannot be determined independently.  However, since 
CS~22949--037 is relatively warm, little CO is formed, and the 
abundance of C is not greatly dependent on the O abundance.  
Nevertheless, we have determined the C, N and O abundances by 
successive iterations and, in particular, the final iteration has been 
performed with a model that takes the observed anomalous abundances of 
these species into account in a self-consistent manner, notably in the 
opacity calculations.\\

\noindent {\em Carbon and nitrogen}\\

\begin {figure}
\resizebox {\hsize}{!}
{\includegraphics {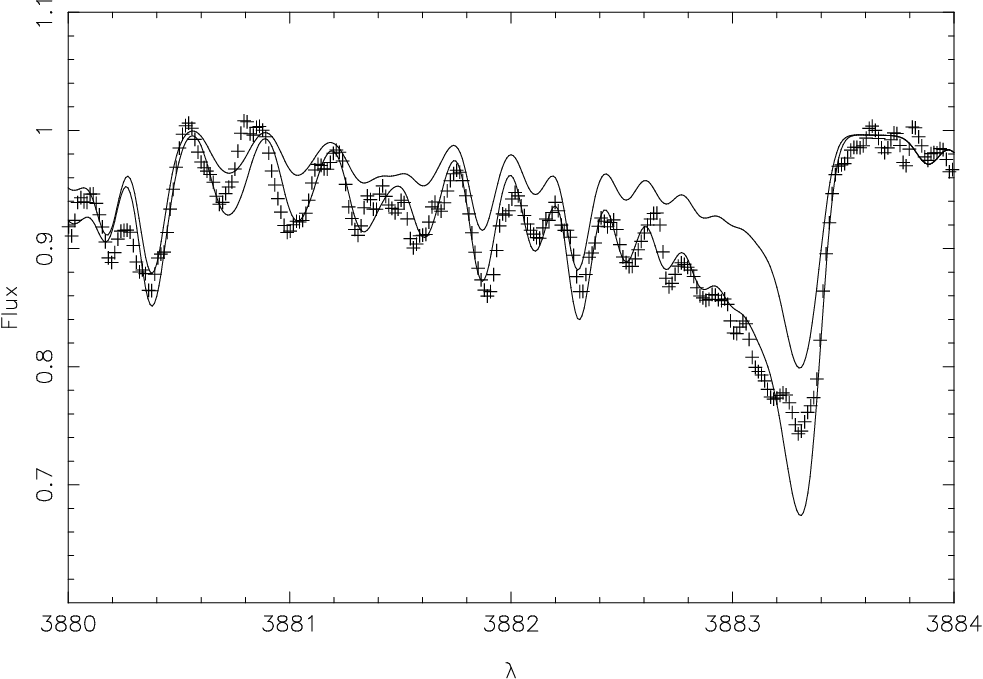}} 
\caption {Comparison of the observed spectrum (crosses) and synthetic 
spectra (thin lines) computed for [C/H] $= -2.80$, and [N/Fe] $= $2.56 
and +2.26., respectively }
\label {CNband}
\end {figure}

The carbon abundance of CS~22949--037 has been deduced from the 
$A^{2}\Delta- X^{2}\Pi$ G band of CH (bandhead at 4323\AA), and the 
nitrogen abundance from the $B^{2}\Sigma - X^{2}\Sigma$ CN violet 
system (bandhead at 3883\AA).  Neither the $A^{3}\Pi_{g} - 
X^{3}\Pi_{u}$ C$_{2}$ Swan band nor the $A^{2}\Pi- X^{2}\Sigma$ red CN 
band, which are often used for abundance determinations, are visible 
in this star.  Line lists for $\mathrm{^{12}CH}$, $\mathrm{^{13}CH}$, 
$\mathrm{^{12}C^{14}N}$, and $\mathrm{^{13}C^{14}N}$ were included in 
the synthesis.  The CN line lists were prepared in a similar manner as 
the TiO line lists of Plez (\cite{pleztio}), using data from Cerny et 
al.  (\cite{cerny}), Kotlar et al.  (\cite{kotlar}), Larsson et al.  
(\cite{larsson}), Bauschlicher et al.  (\cite{bausch}), Ito et al.  
(\cite{ito}), Prasad \& Bernath (\cite{prasada}), Prasad et al.  
(\cite{prasadb}), and Rehfuss et al.  (\cite{rehfuss}).  Programs by 
Kotlar were used to compute wavenumbers of transitions in the red 
bands studied by Kotlar et al.  (\cite{kotlar}).  For CH, the LIFBASE 
program of Luque \& Crosley (\cite{lifbase}) was used to compute line 
positions and $gf$-values.  Excitation energies and isotopic shifts 
were taken from the line list of J\"orgensen et al.  
(\cite{jorgensen}), as LIFBASE only provides line positions for 
$^{12}$CH. This procedure yielded a good fit of the CH lines, except 
for a very few lines which were removed from the list.  Fig.  
\ref{CNband} shows the fit of the CN blue system.

\begin {figure}
\resizebox {\hsize}{!}
{\includegraphics {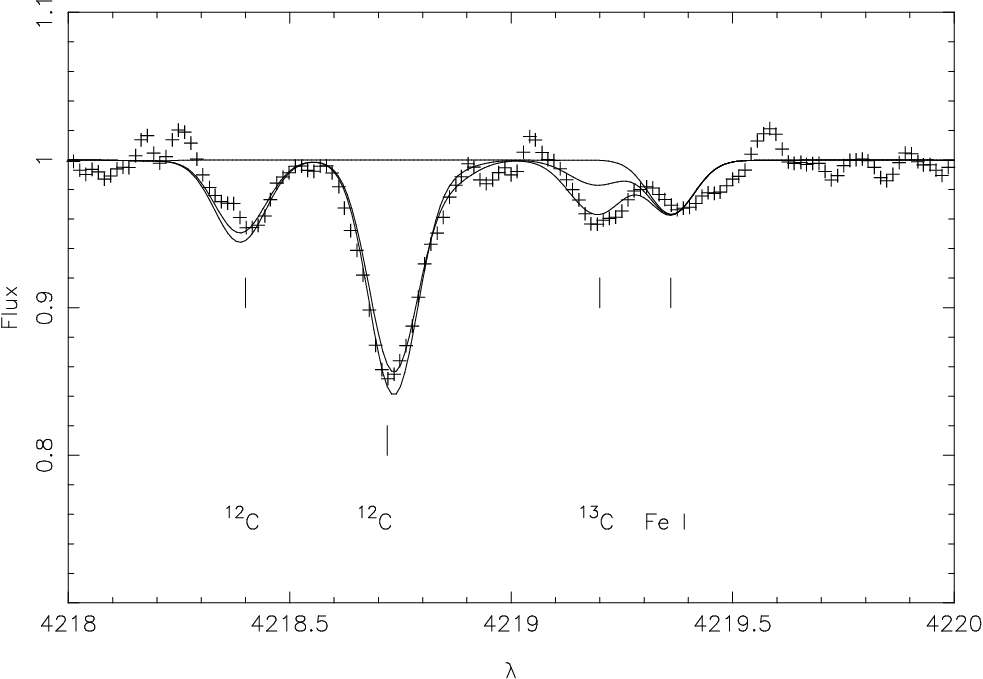}} 
\caption {Comparison of the observed spectrum (crosses) and synthetic profiles 
(thin lines) for $A^{2}\Delta - X^{2}\Pi$ $\mathrm{^{12}CH}$ and 
$\mathrm{^{13}CH}$
lines computed for $^{12}C/^{13}C= 4$ and 10, and with no 
$\mathrm{^{13}C}$.}
\label {C12C13}
\end {figure}

Our analysis confirms the large overabundances of carbon and nitrogen 
in CS~22949--037 ([C/Fe] $= +1.17 \pm 0.1$, [N/Fe] $= +2.56 \pm 0.2$, 
see Table \ref{abon} and Fig.  \ref {CNband}), in good agreement with 
the results of Norris et al.  (\cite {NRB01}).  But as an important 
new result, we have also been able to measure the 
$\mathrm{^{12}C/^{13}C~}$ ratio from $\mathrm{^{13}CH}$ lines of both 
the $A^{2}\Delta - X^{2}\Pi$ and $B^{2}\Sigma - X^{2}\Pi$ systems.  
Using a total of 9 lines of $\mathrm{^{13}CH}$ (6 from the 
$A^{2}\Delta - X^{2}\Pi$ system and 3 from the $B^{2}\Sigma - 
X^{2}\Pi$ system), we find $\mathrm{^{12}C/^{13}C~}= 4.0\pm2$ (Fig.  
\ref{C12C13}).  We note that the wavelengths for the 
$\mathrm{^{13}CH}$ lines arising from the $A^{2}\Delta - X^{2}\Pi$ 
system were systematically $\sim$0.2\AA\ larger than observed, so the 
wavelengths for the whole set of lines was corrected by this amount.

Our derived $\mathrm{^{12}C/^{13}C~}$ ratio is much smaller than that 
found in metal-poor dwarfs: $\mathrm{^{12}C/^{13}C~}= 40$ (Gratton et 
al.  \cite{GSC00}), and is close to the equilibrium isotope ratio 
reached in the CN cycle ($\mathrm{^{12}C/^{13}C~}= 3.0$).  Note also 
that the observed $^{13}$C/$^{14}$N ratio (assuming that the N 
abundance is solely $^{14}$N) is about 0.03, close to the equilibirum 
value of 0.01 (Arnould, Goriely, \& Jorissen \cite{AGJ99}).  We 
conclude that the initial CNO abundances of CS~22949-037 have probably 
been modified by material processed in the equilibrium CN-cycle 
operating in the interior of the star, and later mixed into the 
envelope.

The very large nitrogen abundance of CS~22949--037 ([N/Fe] $= +2.6$) 
has been recently confirmed by Norris et al.  (\cite{NRBA02}) from the 
NH band at 336--337 nm.  Such large overabundances of nitrogen have 
previously been noted in two other carbon-enhanced metal-poor stars, 
CS~22947--028 and CS~22949--034 (Hill et al.  \cite{HBS00}); ([N/Fe] 
$=+1.8$ and [N/Fe] $=+2.3$, respectively), but in these stars the 
carbon overabundance was much larger than in CS~22949--037 ([C/Fe] 
$=+2.0$ and [C/Fe] $=+2.5$, respectively).  Norris et al.  
(\cite{NRB97}) also found a very nitrogen-rich star, CS~22957--027, 
with [N/Fe] $=+2$.  Note, however, that Bonifacio et al.  
(\cite{BMB98}) obtained [N/Fe] $=+1$ for this star, but this discrepancy 
can be probably accounted for by differences in oscillator strenghs adopted for 
the CN band.\\


\noindent{\em Oxygen}\\

The most significant result of this study is that we have detected 
the forbidden [OI] line at 630.031 nm -- the first time that the 
oxygen abundance has been measured in a star as metal deficient as 
CS~22949--037.  The equivalent width of this feature, measured 
directly from the spectrum, is about 6 m\AA. The [OI] line occurs in a 
wavelength range plagued by telluric bands of O$_2$, but the radial 
velocity of CS~22949-037 shifts the oxygen line to a location that is 
far away from the strongest telluric lines in all our spectra.  
Moreover, the position of the [OI] line relative to the telluric lines 
is different in the spectra obtained in August 2000 and September 2001 
(the heliocentric correction varies from 3 to 16 km.s$^{-1}$), which 
provides valuable redundancy in our analysis.

\begin {figure}
\resizebox{8.0cm}{5.7cm}
{\includegraphics {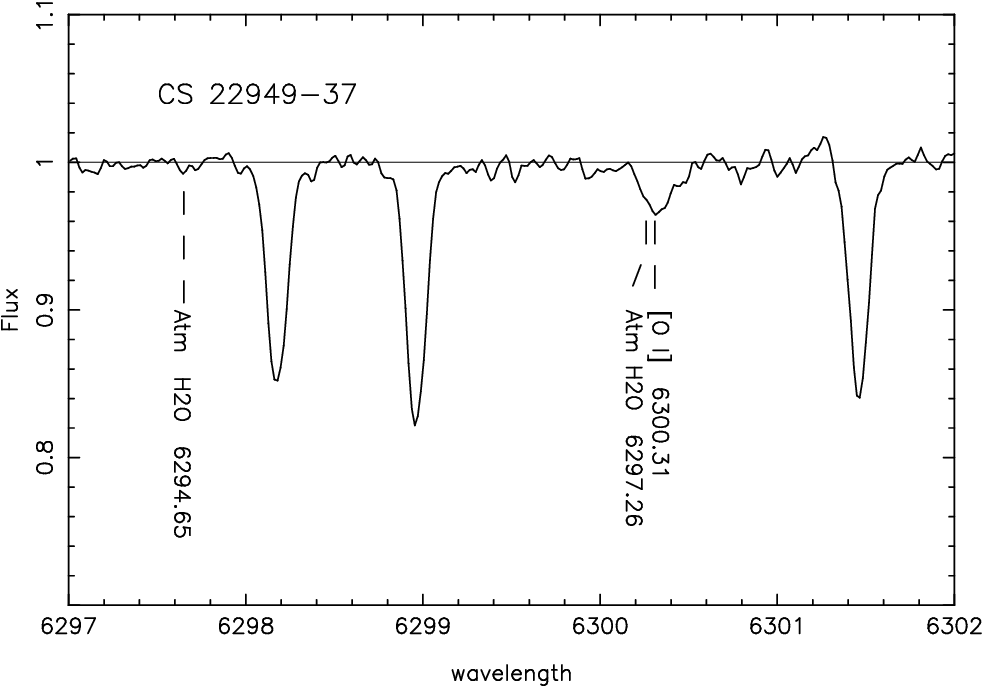}} 
\caption {Mean spectrum of CS~22949--037 from August 2000. The two 
telluric
H$_{2}$O lines are indicated (shifted in wavelength by about 3 \AA\ 
due to the
radial velocity of the star).}
\label {grafoxy}
\end {figure}

\begin {figure}
\resizebox{8.0cm}{5.7cm}
{\includegraphics {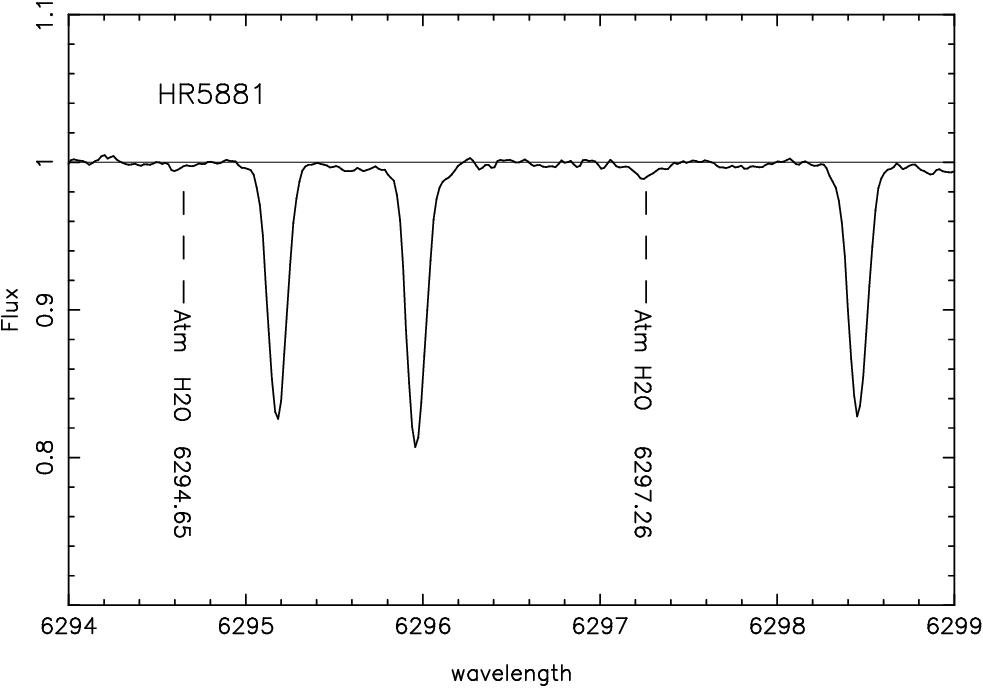}} 
\caption {Mean spectrum of the comparison star HR~5881 (A0~V, 
$v\sin$\,i$ = 87$
km.s$^{-1}$), observed just before CS~22949--037. }
\label {comparoxy}
\end {figure}

\begin {figure}
\begin{center}
\resizebox {3.6cm}{4.5cm} 
{\includegraphics {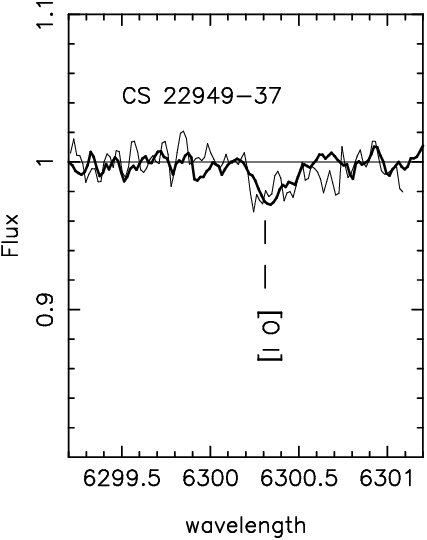}} 
\caption {The spectrum of CS~22949--037 divided by that of HR~5881 to 
eliminate
the telluric lines (heavy line: mean of the three spectra obtained in 
August 2000 spectrum; thin line: September 2001
spectrum). The scale of both axes is the same as in 
Fig.~\ref{grafoxy}. The
measured equivalent width of the corrected [OI] line is 5 m\AA.} 
\label {oxydiv}
\end{center}
\end {figure}

In the August 2000 spectra a weak telluric H$_{2}$O line 
($\lambda=629.726$ nm) is superimposed on the stellar [OI] line (Fig.  
\ref{grafoxy}).  We have accounted for this H$_{2}$O line in two 
different ways.  First, we have estimated its intensity using that of 
another line from the same molecular band system (R1 113), a feature 
observed at 629.465 nm, which should be twice as strong as 
the line at 629.726 nm.  Secondly, we have observed the spectrum of a 
blue star just before that of CS~22949--037, and at about the same 
airmass.  Fig.  \ref{comparoxy} shows the spectrum of the comparison 
star, and Fig.  \ref{oxydiv} the result of dividing our spectra of 
CS~22949--037 by it.

From the telluric line at 629.465 nm in the spectrum of CS~22949--037 
(Fig.  \ref{grafoxy}), we estimate the equivalent width of the line at 
629.726 nm to be $\sim$1.5 m\AA. The equivalent width of the [OI] line 
itself should thus be about 4.5$\pm$1.5~m\AA. From Fig.  \ref 
{oxydiv}, the equivalent width of the OI line is 5~m\AA\,.

In our September 2001 spectrum of CS~22949--037 (Fig.  \ref{oxydiv}) 
the telluric H$_{2}$O line falls outside the region of the stellar 
[OI] line, but the S/N ratio of that spectrum is much lower (Table 
\ref{T-log}).  The measured equivalent width of the [OI] line from 
this spectrum is 5$\pm$2 m\AA, again in good agreement with the 
previous result.  Our final value for the equivalent width of the [OI] 
line at 630.031 nm is then 5.0$\pm$1.0 m\AA, corresponding to [O/H] 
$=-1.98\pm 0.1$ and [O/Fe] $=+1.99 \pm 0.1$ .

\begin {figure}
\begin{center}
\resizebox  {\hsize}{!}
{\includegraphics {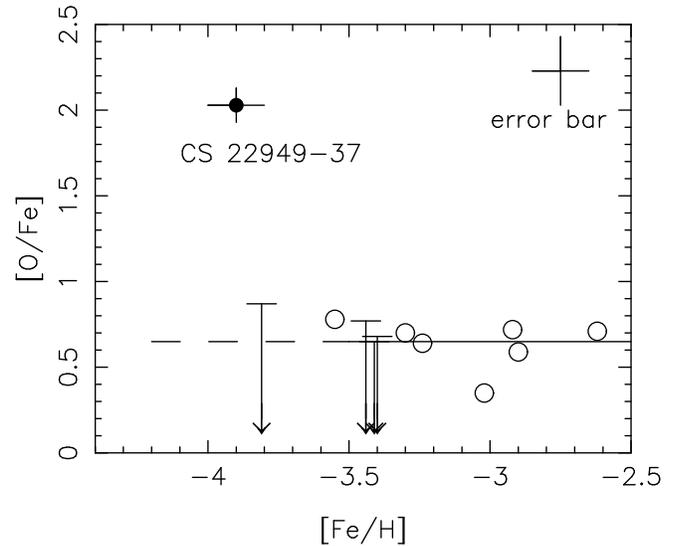} }
\caption {Oxygen abundance from the [O I] forbidden line in extremely
metal-poor stars: CS~22949--037 (black dot, this paper); measured 
values (open circles) and upper limits (arrows) in an extended sample 
(paper in preparation).  At low metallicities, the ratio [O/Fe] does 
not seem to depend on [Fe/H], and is about +0.65.  CS~22949--037 is 
thus very oxygen rich compared to "normal" extremely metal-poor stars.  
The error bar in the upper right corner is the typical uncertainty 
associated with the abundance determination in the sample.  
CS~22949--037 is plotted with its own associated
uncertainties.}
\label {Ompoorst}
\end{center}
\end {figure}

According to Kiselman (\cite {Ki01}) and Lambert(\cite {Lam02}), the 
[OI] feature is not significantly affected by non-LTE effects because 
(a) the line is weak, (b) the transition is a forbidden one (with 
collisional rates largely dominating over radiative rates), and (c) 
the upper level is collisionally excited.  Oxygen abundances derived 
from the [OI] line are therefore less prone to systematic errors, but 
the line is very weak in metal-poor stars, hence high resolution and 
S/N ratio are both required.

Our result that [O/Fe] $\approx +2.0$ in CS~22949--037, the most 
metal-poor star with a measured O abundance, raises the question 
whether such a large overabundance of O is representative of the most 
metal-poor stars in general.  This would be an argument in favour of a 
continued increase of [O/Fe] at the lowest metallicities (e.g., 
Israelian et al.  \cite{IR01}, \cite {IRetal01}).

Our VLT programme includes a sample of XMP giants which were observed 
and analysed in {\em exactly the same way} as CS~22949--037 (Depagne 
et al., in preparation).  To provide a meaningful comparison to 
CS~22949--037, the oxygen abundances derived for this sample (from the 
[O I] line) are shown in Fig.  \ref{Ompoorst}.  We stress once again 
that these abundance measurements have been obtained using exactly the 
same analysis as described in this paper, applied to 11 stars with 
effective temperatures between 4700K and 4900K, surface gravities 
between 0.8 to 1.8 dex, and metallicities in the range $-2.6$ to 
$-3.8$.  The comparison between the oxygen abundance in CS~22949--037 
and the rest of the sample is therefore straighforward, free from 
systematics arising from the method itself (e.g., determination of 
stellar parameters, atmospheric models, choice of oxygen indicator).  
In the metallicity range $-3.5 \le {\rm [Fe/H]} \le -2.5$ we find a 
mean [O/Fe]$\approx +0.65$, with surprisingly little scatter.  We 
therefore expect that at [Fe/H] $=-4.0$, the mean oxygen abundance 
should also be around [O/Fe]$\approx +0.65$, as is also hinted at by 
the upper limit obtained on the [Fe/H] = $-3.8$ giant plotted in Fig.  
\ref{Ompoorst}.  (This point will be discussed in detail, and with a 
larger sample of stars, in a subsequent paper in this series).  Note 
that CS~22949--037 is the only one of the giants in our program with 
[Fe/H] $\approx -4.0$ in which we {\em could} detect the [OI] line: 
for [O/Fe] $=+0.65$ the predicted equivalent width is $\approx$ 0.2 
m\AA, well below the normal detection threshold at this wavelength 
(0.5\---1.0 m\AA, depending on the S/N of the spectra).

We thus conclude that the O abundance in CS~22949--037 is {\em not} 
typical for XMP stars; [O/Fe] appears to be $\sim 1.3$ dex higher than 
the expected abundance ratio for stars of this low metallicity (Fig.  
\ref{Ompoorst}).

We return to the discussion of the origin of these remarkable CNO 
abundances in section \ref{models}.

\subsubsection{The $\alpha$ elements Mg, Si, Ca, and 
Ti}\label{MgSiCa} 

In Fig.  \ref{limet} we compare the light-element abundances of 
CS~22949--037 with those of the well-established XMP giants (the 
``classical'' sample) studied by Norris et al.  (\cite{NRB01}).

The Si abundance in CS~22949--037 has been determined from two lines 
at 390.553 nm and 410.274 nm.  The first is severely blended by a CH 
feature, while the latter falls in the wing of the H$_{\delta}$ line.  
These blends have been taken into account in the analysis, and both 
lines yield similar abundances.  The 869.5 nm and 866.8 nm lines of 
sulphur are not detected, and yield a fairly mild upper limit of 
[S/Fe]$\leq +1.78$.

The even-Z ($\alpha$-) elements Mg, Si, Ca, and Ti are expected to be 
mainly produced during hydrostatic burning in stars, and are generally 
observed to be mildly enhanced in metal-poor halo stars.  It is thus 
remarkable that, in CS~22949--037, the magnitude of the 
$\alpha$-enhancement decreases with the atomic number of the element: 
[Mg/Fe] is far greater than normal, while the enhancement of the 
heaviest $\alpha$-elements, like Ca and Ti, is practically the same as 
in the classical metal-poor sample (a point noted as well by Norris et 
al. \cite{NRB01}).  The [Si/Fe] ratio has an intermediate value.

\subsubsection{The light odd-Z elements Na, Al, and K}\label{NaAlK}

The abundances of the odd-Z elements Na, Al, and K in XMP stars are 
all derived from resonance lines that are sensitive to non-LTE 
effects.  The Al abundance is based on the resonance doublet at 394.4 
and 396.2 nm.  Due to the high resolution and S/N of our spectra, both 
lines can be used, and the CH contribution to the Al I 394.4 nm line 
is easily taken into account.  The Na I D lines are used for the Na 
abundance determination, while the K resonance doublet is not detected 
in CS~22949--037.

\begin {figure} 
\resizebox {\hsize}{!} 
{\includegraphics {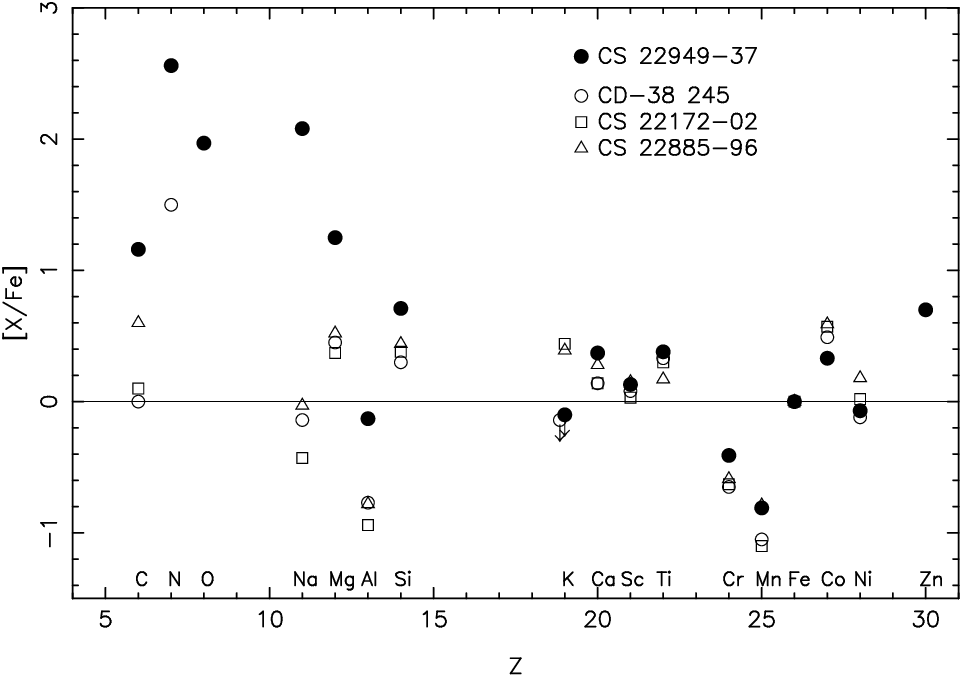}} 
\caption {Abundance of the elements from C to Zn in CS~22949--037 
(filled circles) and in classical XMP stars with [Fe/H]$\approx -4$ 
(open symbols).  N, Mg, and Al are strongly enhanced in CS~22949--037, 
while the behaviour of the subsequent elements is normal.}
\label {limet} 
\end {figure}

Derived [Na/Fe] and [Al/Fe] ratios are usually underabundant in XMP 
field stars (Fig.  \ref{limet}), while we find [K/Fe] to be generally 
overabundant in the classical XMP sample (Depagne et al., in 
preparation), in agreement with Takeda et al.  (\cite {TZC02}).  As 
found already by McWilliam et al.  (\cite{MPS95}), Na is {\em strongly 
enhanced} in CS~22949--037 ([Na/Fe] $=+2.08$), while Al is less 
deficient than normal: [Al/Fe] $=-0.13$ in CS~22949--037, while the 
mean value for the comparison sample is [Al/Fe] $=-0.8$.  The K 
doublet is undetected in both CS~22949--037 and CD --38~245, 
corresponding to upper limits of [K/Fe]$\leq -0.1$ and [K/Fe]$\leq 
-0.14$, respectively.

Several authors have pointed out that, in LTE analyses, the Na 
abundance may be underestimated (Baum\"uller et al.  \cite{BBG98}), 
and the Al and K abundances overestimated (Baum\"uller \& Gehren \cite 
{BG97}; Ivanova \& Shimanskii \cite{IS00}; Norris et al.  \cite{NRB01}).  Following Baum\"uller et al.  (\cite{BBG98}) and 
Baum\"uller \& Gehren (\cite{BG97}), for dwarfs with [Fe/H] $=-3.0$, 
LTE analysis leads to an offset of $\Delta \mathrm{[Al/H] \approx 
-0.65}$ and $\Delta \mathrm{[Na/H] \approx 0.6}$.  Under the 
hypothesis of LTE, Al and Na behave similarly in dwarfs and giants, 
and as a first approximation we can therefore assume that the 
correction is the same for giants as for dwarfs (Norris et 
al.\cite{NRB01}).

\begin {figure}
\resizebox {\hsize}{!}
{\includegraphics {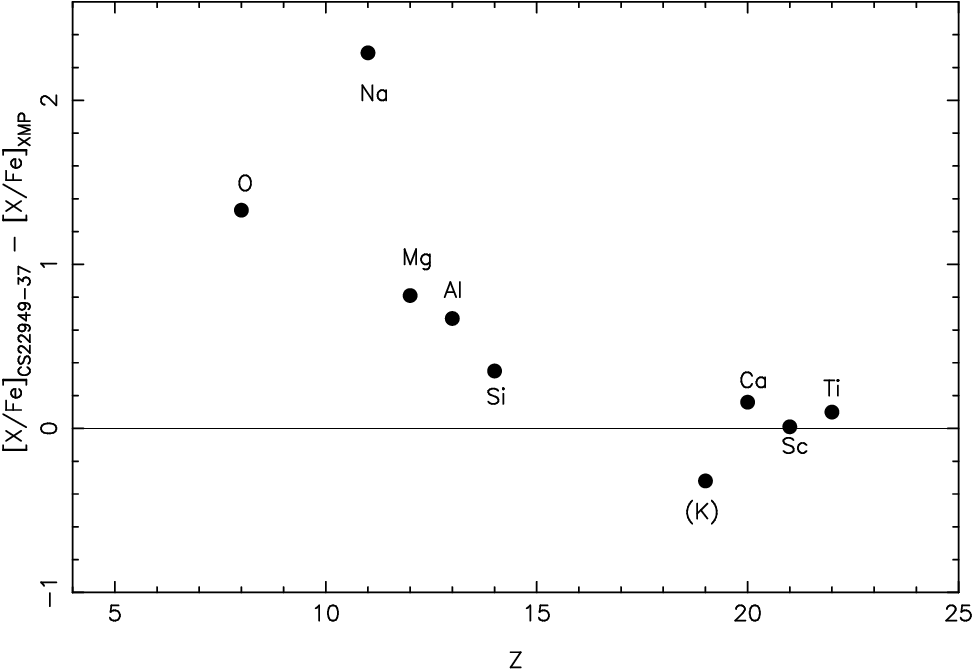}} 
\caption {$\mathrm{[X/Fe]_{22949-037} - [X/Fe]_{XMP}}$, where ``XMP''
represents the mean abundances of the light elements of the three XMP 
giants CD~--38$^{\circ}$245, CS~22172--02, and CS~22885--96.}
\label {limet2}
\end {figure}

However, the atmospheric parameters of all the stars shown in 
Fig.~\ref{limet} are very similar ($\mathrm{4850<T_{eff}<5050}$K, 
$\mathrm{1.7<log~g<2}$, and $\mathrm{[Fe/H]\approx -4}$), so the NLTE 
effects must also be very similar for all four stars.  Accordingly, 
the difference between the Na, Al, and K abundances in CS~22949--037 
and in the classical metal-poor sample must be real and independent of 
the non-LTE effects.  Fig.~\ref {limet2} shows the differences between 
the light-element abundances in CS~22949--037 and the mean of the 
three XMP giants CD--38~245, CS~22172--02, and CS~22885--96.  For 
potassium, the upper limit derived above for CD--38~245 was taken as 
the best approximation to the K abundance of this star when forming 
the mean.  Fig.~\ref {limet2} highlights the dramatic decrease in the 
enhancement of the light elements from Na through Si in CS~22949--037; 
beyond silicon, the abundance ratios in CS~22949--037 are similar to 
those in other XMP stars.

\subsubsection { The Si-burning elements}\label{Sielts}

The distribution of the abundances of the Si-burning elements in 
CS~22949--037 is different from that observed in the Sun, but is 
rather similar to the distribution observed in three other very 
metal-poor stars (see Fig.  \ref{limet}).  Furthermore, this pattern 
is rather well represented by the Z35C zero-metal supernova yields 
obtained by Woosley \& Weaver (\cite{WW95}), as modified by 
Woosley and Heger (model Z35Z, A. Heger, private communication).  
Note that Cr 
and Mn are underabundant while Co and Zn are overabundant.  The 
present analysis of CS~22949-037 is the first case in which a Zn 
abundance has been derived in such an extremely metal-poor star.  We 
find that [Zn/Fe] $= +0.7\pm0.1$, in agreement with the increasing 
trend suggested in other very metal-poor stars (e.g., Primas et al.  
\cite{PRI00}; Blake et al.  \cite{BNRB01}), but of course it should be 
kept in mind that this star may not reflect the general behavior of 
the most metal-deficient stars.

\subsubsection{The neutron-capture elements}\label{subseq-neutron}

CS~22949--037 is a carbon-rich XMP giant, and the neutron-capture 
elements are often (though not always) enhanced in such stars (e.g., 
Hill et al.  \cite{HBS00}).  Sr is indeed enhanced in CS~22949--037 
([Sr/Fe] $=+0.33$), but the [Y/Fe] ratio is about solar ([Y/Fe] $= 
-0.07$).  Ba is underabundant relative to iron ([Ba/Fe] $=-0.58$) while 
Sm and Eu are not detected at all ([Eu/Fe] $\leq 0.04$).  In seeking an 
explanation for the origin of this pattern, we compare the 
distribution of heavy elements in CS~22949--037 to three well-studied 
groups of stars: (i) the classical XMP sample; (ii) the so-called CH 
stars, a well-known class of carbon-rich metal-poor stars, and (iii) 
the new class of mildly carbon-rich metal-poor stars without neutron- 
capture excess (Aoki et al.  \cite{ANRBA02}).  We have excluded the 
r-process enhanced XMP star CS~22892-052, as its mild carbon 
enhancement ([C/Fe] $\approx +1$) appears to be unique among the 
presently known examples of this class.\\

{\em{(i)} The classical XMP stars}\\
In the three classical XMP giants (Norris et al.  \cite{NRB01}), both 
Sr and Ba are very deficient with respect to iron, and by about the 
same factor: [Sr/Fe]$_{mean}= -1.16$, [Ba/Fe]$_{mean}= -1.15$ and 
[Ba/Sr]$_{mean}\approx 0.0$.  In CS~22949--037, Sr is actually 
enhanced and [Ba/Sr] $= -0.91$, which is a rather different pattern.\\

\noindent{\em(ii) The CH stars }\\
As a group, the CH stars are moderately metal-poor ([Fe/H] $= -1.5$) 
yet strongly enriched in neutron-capture elements; most of them 
presumably formed as the result of the {\em s}-process inside an AGB 
star (Vanture \cite{Van92}).  The CH stars are members of long-period 
binary systems with orbital characteristics consistent with the 
presence of a fainter companion, and it is generally assumed that the 
abundance anomalies in these stars are the result of mass transfer 
from the AGB companion, which has now evolved into a white dwarf.  At 
least one very metal-poor star is known to share many of these 
properties (LP 625-44; Norris et al.  \cite{NRB97}), and it too is a 
member of a binary system.  More recently, Aoki et al. (\cite{ANRBA00}) 
have confirmed the earlier work that suggested the level of 
[Sr/Ba] increases with atomic number ( at contrast 
with our star), as expected from an $s$-process at high neutron 
exposure (e.g. Wallerstein et al.  \cite {WIP97}; Gallino et al.  
\cite{GAB98}).  For example, in some of these very metal-poor CH 
stars, $^{208}$Pb is extremely overabundant (Van Eck et al.  \cite 
{EGJ01}); this element is not even detected in our star.  In 
CS~22949--037 the situation is the opposite: the heavy-element 
abundances {\em decrease} with atomic number and even turn into a 
deficit ([Sr/Fe] $= +0.33$, [Y/Fe] $=-0.07$, and [Ba/Fe] $=-0.58$).  
Accordingly, the neutron-exposure processes in CS~22949--037 and the 
CH stars appear to have been completely different.  Moreover, there is 
no indication that CS~22949--037 belongs to a binary system (cf.  
Table \ref {T-log} and section 2), but since the CH stars are 
generally long-period, low-amplitude binaries we cannot
exclude that our star has been enriched by a companion. Accurate longer-term 
monitoring of the radial velocity of CS~22949--037 will be needed to 
settle this point.\\

\noindent {\em(iii) The carbon-rich metal-poor stars without 
neutron-capture excess}\\
Norris, Ryan \& Beers (\cite {NRB97b}), Bonifacio et al.  (\cite{BMB98}) and Aoki et al. (\cite{ANRBA02}) 
observed five very metal-poor, carbon-rich giants ($\mathrm{-3.4 
<[Fe/H]< -2.7}$) without neutron-capture element excess.  Like 
CS~22949--037, these stars exhibit carbon excesses of [C/Fe] $\approx 
+1$, but $\mathrm{^{12}C/^{13}C~}\approx 10$.  the nitrogen excesses 
in these stars is much smaller ($\mathrm{0.0 <[N/Fe]< 1.2}$), and the 
neutron-capture abundances are about the same as in normal halo giants 
with [Fe/H]$\approx -3$.  Sr and Ba are underabundant, but the [Ba/Sr] 
ratio is close to solar, again completely unlike CS~22949--037.  Aoki 
et al.  (\cite{ANRBA02}) have suggested that this class of stars could 
be the result of helium shell flashes near the base of the AGB in very 
low-metallicity, low-mass stars, but this hypothesis is not yet 
confirmed.

In summary, the detailed abundance patterns in CS~22949--037 appear to 
require a different origin from that of the currently known groups of 
carbon-rich XMP giants.  This point is discussed further in the 
next section.

\section{Comparison of the abundance pattern of CS 22949--037 with 
various theoretical studies}\label{models}

This early generation star is not the first one to display an 
abundance pattern that is not easily accounted for by 
standard SN II nucleosynthesis computations. The HK survey has 
found several  very iron-poor stars with large
abundances of C, and  N ( e.g., Hill et al. (\cite{HBS00})).
 Fig. \ref {limet} compares the abundances in CS 22949--037 with those 
of 3 classical metal-poor stars, which are passably explained by 
current SN II nucleosynthesis (Tsujimoto et al. 1995; Woosley \& Weaver
1995), although no 
  nitrogen (which is 
by-passed in a pure helium core) is predicted in our star and in 
CD--38$^{\circ}$245, at contrast with the observations. 
Fig. \ref{limet2}  gives the 
ratios of the abundances in CS 22949--037 to the mean of the 3 stars. 
Very clearly the major feature is a large 
relative overabundance with respect  to iron of the light elements 
C, N, O, Na, and Mg,  declining to almost insignificance
at Si, and none for $Z > 15$, as already noted in Norris et al. (\cite{NRB01}). Qualitalively, something very similar is 
occurring in the model Z35B of Woosley \& Weaver
(\cite{WW95}) which, because of insufficient explosion energy and 
partial "fallback",  expels only C, O, Ne, Na, Mg, Al, a 
very small quantity of Si, and nothing heavier. Below we discuss several attempt to refine this idea.

Another path was followed by Norris et al. (\cite{NRBA02}) for 
explaining CS 22949--037: the pair-instability hypernova  yields 
(see Fryer, Woosley \& Heger \cite{FWH01}, and Heger \& Woosley 
\cite{HW02}).  Here one important ingredient is the mixing of some of 
the carbon in the helium core with proton-rich material, producing a 
large amount of primary nitrogen.  However, the other yields of pair-instability
 supernovae have some features which poorly fit the more 
complete pattern we have obtained here for CS 22949--037.  In 
particular, they show a larger odd/even effect than the one seen in the 
star, and a small [Zn/Fe] , in contrast to the observed value of [Zn/Fe]$= +0.7$.  So, it seems that, 
if the idea of primary nitrogen production by 
mixing must be retained, the case for pair-instability hypernovae is 
not attractive.  

A large body of other theoretical work is relevant to the nucleosynthesis 
in very low-metallicity stars, and we make no attemps to fully sumarize 
previous results in the present papaer. However, a few recent ideas are worth 
keeping in mind. For example, Umeda \& 
Nomoto (\cite{UN02}) have tried to explain the [Zn/Fe] $\approx 0.5$ 
found at very low metallicity.  
Their conclusion is that the solution is a combination of a proper 
mass cut, followed by mixing between the initial mass cut and the top 
of the incomplete Si-burning region, followed by a fallback of most of 
the Si-burning region.  In order to produce the usual [O/Fe] value and 
[Zn/Fe] $\approx 0.5$, it is necessary to have a progenitor mass of 25 
or 30 M$_{\sun} $, {\it and} an energetic explosion of 10 to 
30$\times10^{51}$ ergs.

Chieffi  \& Limongi (\cite {CL02}) have explored the possibility of 
adjusting the free parameters in a single SN II event to fit the abundances of five individual very 
metal-poor stars ( Norris et al. 
\cite{NRB01}, including CS 22949--037).  
Although  in the end they discard CS 22949--037, they note that, except for 
the overabundance of C to Mg, the star
is very similar to the other stars of the sample, and that the high 
[Co/Fe] value is apparently well explained in all C-rich 
stars by their computed yields.

\begin {figure} 
\resizebox {\hsize}{!} {\includegraphics {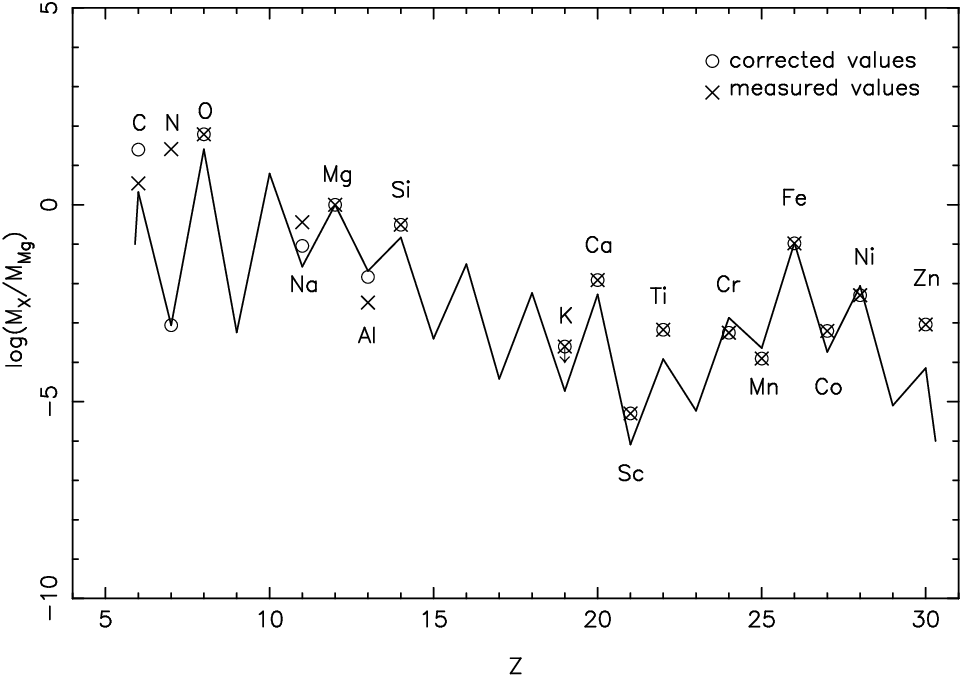}} 
\caption { The logarithmic mass ratio of elements X to Mg (log 
$M_{X}/M_{Mg}$) compared with those predicted by the zero 
heavy-element supernova model Z35Z (Woosley \& Weaver \cite{WW95}, as 
recently modified by Heger and Woosley).  The measured abundances in 
CS~22949--037 have been corrected for NLTE effects (Na, Al), or for 
internal mixing in the star (C, N).  The open circles represent the 
assumed initial abundances of the elements, while the crosses show the 
atmospheric abundances as derived in LTE. For K, only an upper limit 
is available.} 
\label {wooheg2} 
\end {figure}

Finally, we come back to the "fallback" explanation for the high, 
C,N,O, and Na abundances, which make this [Fe/H] =-4 star a Z$= 0.01 
$Z$_{\sun}$ star.  An unpublished result (model Z35Z of Woosley \& 
Heger, in preparation) was kindly communicated to us as a variant of 
the already cited model Z35C. This model has a slightly larger amount 
of fallback, and includes hydrodynamical mixing in the explosion.  It 
shows a fairly good  fit with our observations (crosses in Fig.  \ref {wooheg2}), except for Al and Na, which have to be corrected for non-LTE effects, and for N, 
which is not expected to be formed in the Z35Z 
model.  To improve this fit we corrected for the non-LTE effects on Na and 
Al (see section 3.2.3), and we supposed (open circles in Fig. \ref {wooheg2}) that the observed 
abundance of nitrogen was the result of a transformation of carbon 
into nitrogen through the CN cycle (in the star itself or in its 
progenitor).  After these corrections the agreement is much better. The discrepancy about the Zn abundance is probably curable  (Umeda \&Nomoto \cite {UN02}) as explained here above. 

At this point we must mention that rotation may  be a source for 
mixing and CN processing (see Meynet \& Maeder 2002),
and that other non-standard mixing mechanisms have been investigated 
along the RGB, which may have altered 
the $^{12}$C/$^{13}$C ratio  and the C/N ratio in CS 22949--037 itself 
(Charbonnel \cite{C95}).

The computation of the supernova yields does not contain predictions 
for the neutron-capture elements.  Generally speaking, these elements 
are not easily built in zero-metal supernovae (like Z35Z), nor in 
zero-metal very massive objects, owing to an inefficient neutron flux,
a lack of neutron seeds or both.  The main phenomenon observed in 
CS~22949--37 is the very rapid decline of the abundance of these 
elements with the atomic number.  Such a decline is not observed in 
other very metal-poor stars (see section \ref {subseq-neutron}), and it 
suggests an unusually "truncated" neutron exposure (very short 
relative to the neutron flux).

In summary, it appears that SNe II of mass near 30 
M$_{\sun}$ , either primordial or of very low metallicity,
offer good  prospects for explaining stars like CS 22949--037. 
Enough ingredients are available.  They have still to be 
assembled in the most economic way.

\section {Conclusions}

Bringing the light collecting power, resolution, and extended wavelength coverage of 
UVES/VLT to bear on the abundance analysis of CS~22949--37 has provided important new
 results as well as refined the results of previous analyses. For the first time in a 
 such metal poor star, we have measured the forbidden OI line at 630.0 nm: we found
 [O/Fe]$= +1.97 $.  We found a mild carbon enhancement [C/Fe]$=+1.17$ and a very
 low $\mathrm{^{12}C/^{13}C~}= 4\pm2$ ratio, close to the equilibrium value
 The elemental abundances of the extremely metal-poor stars 
 CS~22949--037 are very unusual. The strong enhancement of oxygen (not typical 
for this very low metallicity) could be explained by pair-instability 
supernovae, but the strong odd-even effect predicted in these models, 
which is not observed, rules out these very massive objects.
The enhancement of O and C is explained by models of zero-metallicity (or 
very metal-poor) core-collapse supernovae. The observed enhancement of 
N has to be explained by the CN processing of C, in the supernova or 
in the star itself (or its companion, if binary). The fair agreement between the observed
elemental abundances in CS~22949--037 and those predicted in the Z35Z
model of Woosley and Heger (private communication) suggests that the most likely interpretation
is that this star exhibits the ejecta of a single core-collapse supernova.
However, more complex scenarios, in which the combined ejecta of several progenitors
are responsible, are not excluded by the present data. Clearly, 
interpretation of the exceptional pattern of the neutron-capture elements
in CS~22949--037 merits further study. The identification of other
extremely metal-poor stars that exhibit similar patterns would be most illuminating.

\begin {acknowledgements} We thank A. Heger, S. Woosley, I. Baraffe, 
A. Chieffi and L. Limongi for useful discussions pertaining to the 
interpretation of the elemental abundance patterns in CS~22949-037.  
In particular A. Heger and S. Woosley for kindly providing the 
modified supernova yields and A. Chieffi and L. Limongi for sending 
a copy of their paper in  advance of publication their paper.  We thank N. 
Jacquinet-Husson for helping to extract accurate wavelengths of the 
telluric lines from the GEISA database.  We also thank the referee, J. 
Cowan, for useful comments to improve the paper.  J.A. and B.N. 
received partial support for this work from the Carlsberg foundation 
and the Danish Natural Science Research Council.  T.B. acknowledges 
partial support of this work from the U.S. National Science 
Foundation, in the form of grants AST 00-98549 and AST 00-98508.  
\end {acknowledgements}

\newpage
\appendix

\begin {table}[h]
{\bf \large \center{Appendix}}\\  Line list and atomic data.\\ We list in this table all the lines that were used to derive abundance.\\
\caption {Linelist, equivalent width, and abundance for the elements in CS~22949--937}
\begin {tabular}{cccccc}
\hline
Element&       &         &         &           & $<\log \epsilon>$\\
\hline
&  lambda  &  log(gf)& W (m\AA\ ) & $\log \epsilon$   & \\
\hline
O1&          &         &         &        &  \bf {6.84} \\
\hline
& 6300.304 & -9.750  &   5.0 &    6.84  &\\
\hline
Na1&          &         &       &	    &  \bf {3.80} \\
\hline
& 5889.951 &  0.110  & 151.6 &    3.82  & \\
& 5895.924 & -0.190  & 132.9 &    3.77  & \\
\hline
Mg1&          &         &       &	    &  \bf {5.17} \\
\hline
& 3829.355 & -0.210  & 156.1 &    5.19  & \\
& 3832.304 &  0.150  & 185.2 &    5.12  & \\
& 3838.290 &  0.420  & 202.7 &    4.85  & \\
& 4167.271 & -1.000  &  46.3 &    5.54  & \\
& 4571.096 & -5.390  &  52.9 &    5.07  & \\
& 5172.684 & -0.380  & 176.1 &    5.22  & \\
& 5183.604 & -0.160  & 199.4 &    5.23  & \\
& 5528.405 & -0.340  &  62.6 &    5.08  & \\
\hline
Al1 &          &         &       &      &  \bf {2.34} \\
\hline
& 3944.006 & -0.640  &  71.5 &    2.36  & \\
& 3961.520 & -0.340  &  84.1 &    2.32  & \\
\hline
Si1  &          &         &       &           &  \bf {4.05} \\
\hline
& 4102.936 & -2.700  &  16.4 &    4.26  & \\
\hline
Ca1  &          &         &       &           &  \bf {2.73} \\
\hline
& 4226.728 &  0.240  & 112.5 &    2.67  & \\
& 4283.011 & -0.220  &  15.4 &    3.12  & \\
& 4318.652 & -0.210  &   6.1 &    2.67  & \\
& 4454.779 &  0.260  &  19.4 &    2.77  & \\
& 5588.749 &  0.210  &   4.1 &    2.70  & \\
& 5857.451 &  0.230  &   1.0 &    2.51  & \\
& 6102.723 & -0.790  &   2.3 &    2.70  & \\
& 6122.217 & -0.320  &   8.1 &    2.81  & \\
& 6162.173 & -0.090  &  11.4 &    2.76  & \\
& 6439.075 &  0.470  &   5.1 &    2.52  & \\
\hline
Sc2  &          &         &       &           &  \bf {-0.70}\\
\hline
& 4246.822 &  0.240  &  62.7 &   -0.72  & \\
& 4314.083 & -0.100  &  37.8 &   -0.47  & \\
& 4400.389 & -0.540  &  11.7 &   -0.71  & \\
& 4415.557 & -0.670  &   8.5 &   -0.75  & \\
& 5031.021 & -0.400  &   2.0 &   -0.84  & \\
\hline
Ti1  &          &         &       &           &  \bf {1.40} \\
\hline
& 3998.636 & -0.060  &  11.9 &    1.35  & \\
& 4533.241 &  0.480  &   7.2 &    1.44  & \\
& 4534.776 &  0.280  &   3.4 &    1.29  & \\
& 4981.731 &  0.500  &   9.0 &    1.48  & \\
& 4991.065 &  0.380  &   6.5 &    1.45  & \\
& 4999.503 &  0.250  &   5.8 &    1.51  & \\
& 5173.743 & -1.120  &   1.8 &    1.38  & \\
& 5192.969 & -1.010  &   1.7 &    1.27  & \\
\hline
Ti2  &          &         &       &           &  \bf {1.41} \\
\hline
& 3759.296 & -0.460  & 117.7 &    2.24  & \\
& 3761.323 &  0.100  & 114.6 &    1.56  & \\
& 3913.468 & -0.530  &  67.0 &    1.54  & \\
& 4012.385 & -1.610  &  31.9 &    1.28  & \\
& 4028.343 & -1.000  &   5.4 &    1.28  & \\
& 4290.219 & -1.120  &  37.4 &    1.55  & \\
& 4337.915 & -1.130  &  35.1 &    1.42  & \\
\end {tabular}
\end {table}

\newpage

\begin {table}[h]
\begin {tabular}{cccccc}
& 4395.033 & -0.660  &  56.9 &    1.34  & \\
& 4395.850 & -2.170  &   2.8 &    1.37  & \\
& 4399.772 & -1.270  &  22.1 &    1.46  & \\
& 4417.719 & -1.430  &  24.5 &    1.59  & \\
& 4443.794 & -0.710  &  51.5 &    1.29  & \\
& 4450.482 & -1.450  &  19.1 &    1.37  & \\
& 5226.543 & -1.290  &   6.10&    1.165 & \\
\hline
Cr1  &          &         &       &           &  \bf {1.29} \\
\hline
& 4254.332 & -0.110  &  38.2 &    1.16  & \\
& 4274.796 & -0.230  &  35.7 &    1.23  & \\
& 4289.716 & -0.360  &  31.8 &    1.28  & \\
& 5206.038 &  0.020  &  10.9 &    1.33  & \\
& 5208.419 &  0.160  &  17.5 &    1.43  & \\
\hline
Mn1  &          &         &       &           &  \bf {0.61} \\
\hline
& 4030.753 & -0.480  &  26.6 &    0.60  & \\
& 4033.062 & -0.620  &  21.2 &    0.61  & \\
\hline
Fe1  &          &         &       &           &  \bf {3.51} \\
\hline
& 3899.700 & -1.530  & 102.2 &    3.84  & \\
& 3920.300 & -1.750  &  95.5 &    3.87  & \\
& 3922.900 & -1.650  & 102.5 &    3.91  & \\
& 4005.200 & -0.610  &  61.5 &    3.45  & \\
& 4045.800 &  0.280  &  99.6 &    3.46  & \\
& 4063.600 &  0.070  &  89.8 &    3.47  & \\
& 4071.700 & -0.020  &  84.1 &    3.46  & \\
& 4076.600 & -0.370  &   6.3 &    3.62  & \\
& 4132.100 & -0.670  &  58.2 &    3.47  & \\
& 4143.900 & -0.460  &  66.2 &    3.37  & \\
& 4147.700 & -2.100  &   7.7 &    3.48  & \\
& 4156.800 & -0.610  &   5.0 &    3.32  & \\
& 4174.900 & -2.970  &   7.4 &    3.67  & \\
& 4181.800 & -0.180  &  10.5 &    3.24  & \\
& 4187.000 & -0.550  &  18.6 &    3.47  & \\
& 4187.800 & -0.550  &  22.1 &    3.53  & \\
& 4191.400 & -0.730  &  13.5 &    3.51  & \\
& 4195.300 & -0.410  &  11.1 &    4.06  & \\
& 4199.100 &  0.250  &  23.9 &    3.49  & \\
& 4202.000 & -0.700  &  60.3 &    3.38  & \\
& 4222.200 & -0.970  &   8.3 &    3.48  & \\
& 4227.400 &  0.230  &  15.1 &    3.57  & \\
& 4233.600 & -0.600  &  19.0 &    3.56  & \\
& 4250.100 & -0.400  &  22.6 &    3.45  & \\
& 4260.500 & -0.020  &  44.2 &    3.44  & \\
& 4271.200 & -0.350  &  32.7 &    3.60  & \\
& 4271.800 & -0.160  &  88.1 &    3.50  & \\
& 4282.400 & -0.820  &  16.0 &    3.33  & \\
& 4325.800 & -0.010  &  91.2 &    3.57  & \\
& 4404.800 & -0.140  &  85.1 &    3.46  & \\
& 4415.100 & -0.610  &  61.8 &    3.44  & \\
& 4447.700 & -1.340  &   6.0 &    3.42  & \\
& 4461.700 & -3.200  &  33.0 &    3.71  & \\
& 4466.600 & -0.600  &   6.0 &    3.37  & \\
& 4494.600 & -1.140  &  11.7 &    3.51  & \\
& 4528.600 & -0.820  &  21.3 &    3.48  & \\
& 4871.300 & -0.360  &  10.7 &    3.43  & \\
& 4872.100 & -0.570  &   7.8 &    3.50  & \\
& 4891.500 & -0.110  &  19.8 &    3.47  & \\
& 4919.000 & -0.340  &  11.0 &    3.42  & \\
& 4920.500 &  0.070  &  23.5 &    3.36  & \\
& 4994.100 & -3.080  &   7.5 &    3.72  & \\
& 5001.900 &  0.010  &   3.1 &    3.61  & \\
& 5041.100 & -3.090  &   4.2 &    3.51  & \\
& 5041.800 & -2.200  &   8.2 &    3.54  & \\
\end {tabular}
\end {table}

\newpage

\begin {table}[h]
\begin {tabular}{cccccc}
& 5049.800 & -1.360  &   5.5 &    3.43  & \\
& 5051.600 & -2.800  &  11.2 &    3.63  & \\
& 5068.800 & -1.040  &   3.9 &    3.70  & \\
& 5110.400 & -3.760  &  15.9 &    3.68  & \\
& 5123.700 & -3.070  &   4.8 &    3.60  & \\
& 5127.400 & -3.310  &   3.0 &    3.52  & \\
& 5166.300 & -4.200  &   7.8 &    3.77  & \\
& 5171.600 & -1.790  &  19.5 &    3.56  & \\
& 5194.900 & -2.090  &   6.9 &    3.42  & \\
& 5232.900 & -0.060  &  15.6 &    3.37  & \\
& 5266.600 & -0.390  &   6.2 &    3.32  & \\
& 5324.200 & -0.240  &   7.2 &    3.48  & \\
& 5328.500 & -1.850  &  15.0 &    3.55  & \\
& 5339.900 & -0.720  &   3.2 &    3.65  & \\
& 5371.500 & -1.650  &  62.6 &    3.63  & \\
& 5383.400 &  0.640  &   2.3 &    3.32  & \\
& 5397.100 & -1.990  &  46.8 &    3.64  & \\
& 5405.800 & -1.840  &  44.8 &    3.53  & \\
& 5429.700 & -1.880  &  48.5 &    3.60  & \\
& 5434.500 & -2.120  &  31.5 &    3.59  & \\
& 5446.900 & -1.910  &  43.8 &    3.58  & \\
& 5455.600 & -2.090  &  33.4 &    3.59  & \\
& 5506.800 & -2.800  &   9.4 &    3.59  & \\
\hline
Fe2  &          &         &       &           &  \bf {3.56} \\
\hline
& 4178.862 & -2.480  &   5.0 &    3.39  & \\
& 4233.172 & -2.000  &  20.2 &    3.59  & \\
& 4416.830 & -2.600  &   4.9 &    3.71  & \\
& 4515.339 & -2.480  &   3.4 &    3.49  & \\
& 4520.224 & -2.610  &   2.8 &    3.50  & \\
& 4555.893 & -2.280  &   7.3 &    3.62  & \\
\hline
Co1  &          &         &       &            &  \bf {1.28} \\
\hline
& 3845.461 &  0.010  &  30.9 &    1.35  & \\
& 3995.302 & -0.220  &  16.7 &    1.20  & \\
& 4118.767 & -0.490  &   7.9 &    1.24  & \\
& 4121.311 & -0.320  &  17.2 &    1.30  & \\
\hline
Ni1  &          &         &       &           &  \bf {2.19} \\
\hline
& 3807.138 & -1.180  &  50.4 &    2.21  & \\
& 3858.292 & -0.970  &  58.9 &    2.17  & \\
& 5476.900 & -0.890  &   6.2 &    2.17  & \\
& 5476.900 & -0.890  &   6.2 &    2.17  & \\
\hline
Zn1  &          &         &       &           &  \bf {1.29 } \\
\hline
 &4822.528 & -0.13   &   3.0 &   1.29   & \\ 
\hline
Sr2  &          &         &       &           &  \bf {-0.72} \\
\hline
& 4077.709 &  0.170  & 109.3 &   -0.67  & \\
& 4215.519 & -0.170  &  95.5 &   -0.75  & \\
\hline
Y2   &          &         &       &           &  \bf {-1.80} \\
\hline
& 3950.352 & -0.490  &	 7.7 &   -1.68  & \\
& 3950.352 & -0.490  &	 7.7 &   -1.68  & \\
& 4883.684 &  0.070  &	 1.6 &   -1.90  & \\
& 5087.416 & -0.170  &	 1.2 &   -1.80  & \\
\hline
Ba2  &          &         &       &           &  \bf {-2.42} \\
\hline
& 4554.029 &  0.170  &  23.1 &   -2.53  & \\
& 4934.076 & -0.150  &  15.7 &   -2.46  & \\
& 5853.668 & -1.010  &   0.7 &   -2.38  & \\
& 6141.713 & -0.070  &   5.6 &   -2.30  & \\
\hline
Sm2  &          &         &        &         &  \bf {-1.82} \\
\hline
& 4537.941 & -0.230  &   1.3 &   -1.82  & \\
\hline
Eu2  &          &         &        &           &  \bf {-3.42} \\
\hline
& 4129.725 &  0.200  &	 1.00 &  -3.41  & \\
\end {tabular}
    
\end {table}
\end {document}